\documentclass{article}

\usepackage{arxiv}

\usepackage[utf8]{inputenc} 
\usepackage[T1]{fontenc}    
\usepackage{hyperref}       
\usepackage{url}            
\usepackage{booktabs}       
\usepackage{amsfonts}       
\usepackage{nicefrac}       
\usepackage{microtype}      
\usepackage{lipsum}		
\usepackage{graphicx}
\usepackage{doi}
\usepackage{amsthm}
\theoremstyle{plain}
\newtheorem{theorem}{Theorem}

\newtheorem{proposition}[theorem]{Proposition}

\theoremstyle{definition}

\newtheorem{example}{Example}

\theoremstyle{remark}
\newtheorem{remark}{Remark}

\title{On synchronization in Kuramoto models on spheres}

\author{Aladin Crnki\' c \\
	Faculty of Technical Engineering\\
	University of Biha\' c\\
	Irfana Ljubijanki\' ca bb., 77000 Biha\' c\\
	Bosnia and Herzegovina\\
	\texttt{aladin.crnkic@unbi.ba} \\
\And
Vladimir Ja\' cimovi\' c \\
	Faculty of Natural Sciences and Mathematics\\
	University of Montenegro\\ 
	Cetinjski put bb., 81000 Podgorica\\ 
	Montenegro\\
	\texttt{vladimirj@ucg.ac.me} \\
	\And
	Marijan Markovi\' c \\
Faculty of Natural Sciences and Mathematics\\
	University of Montenegro\\ 
	Cetinjski put bb., 81000 Podgorica\\ 
	Montenegro\\
	\texttt{marijanm@ucg.ac.me} \\
}

\renewcommand{\shorttitle}{On synchronization in Kuramoto models on spheres}

\hypersetup{
pdftitle={On synchronization in Kuramoto models on spheres},
pdfsubject={-},
pdfauthor={Aladin Crnki\' c, Vladimir Ja\' cimovi\' c, Marijan Markovi\' c},
pdfkeywords={non-Abelian Kuramoto model, sphere, conformal mappings, Poisson kernels},
}

\begin{document}
\maketitle

\begin{abstract}
We analyze two classes of Kuramoto models on spheres that have been introduced in previous studies. Our analysis is restricted to ensembles of identical oscillators with the global coupling. In such a setup, with an additional assumption that the initial distribution of oscillators is uniform on the sphere, one can derive equations for order parameters in closed form.

The rate of synchronization in a real Kuramoto model depends on the dimension of the sphere. Specifically, synchronization is faster on higher-dimensional spheres. On the other side, real order parameter in complex Kuramoto models always satisfies the same ODE, regardless of the dimension.

The derivation of equations for real order parameters in Kuramoto models on spheres is based on recently unveiled connections of these models with geometries of unit balls.

Simulations of the system with several hundreds of oscillators yield perfect fits with the theoretical predictions, that are obtained by solving equations for the order parameter.
\end{abstract}


\section{Introduction}\label{sec:1}
Large populations of coupled oscillators display a fascinating variety of collective phenomena that have been observed and studied for centuries, see \cite{PRK}. The most popular model in the field has been proposed in 1975. by Yoshiki Kuramoto in \cite{Kuramoto}. The state of each oscillator in the Kuramoto model is given by a single phase variable $\varphi_j \in [0,2 \pi]$, while amplitudes are neglected. Therefore, Kuramoto oscillators can be represented by points $z_j = e^{i \varphi_j}$ on the unit circle $S^1$ in the complex plane.

The second crucial assumption incorporated in the model is that the coupling between each pair of oscillators is proportional to the sine of their phase difference. In other words, oscillators are coupled through the first harmonic only and not through higher harmonics.

There is a growing interest in various extensions of the classical Kuramoto model to higher-dimensional manifolds. Such extensions describe populations of coupled generalized "oscillators" whose states are represented by points on some higher-dimensional manifolds, rather than $S^1$. Interest in such extensions is motivated by their relevance in modeling collective dynamical phenomena in networks of interacting agents, but also by relations with some intriguing theories from mathematics and mathematical physics.

Extensions of the classical Kuramoto model to higher dimensions are based on generalizations of the notion of oscillator. One approach is to consider generalized "oscillators" whose states are described by points on a certain Lie group. Then the intrinsic frequencies of oscillators are elements of the corresponding Lie algebra. Each individual oscillator is governed by a linear ODE on a Lie group. Introduction of coupling into the model yields a system of geometric Riccati ODE's on this group, see \cite{Lohe3}. Typical examples of this kind are Kuramoto models on some matrix groups, such as $SU(k)$ or $SO(n)$. Such models are usually referred to as {\it non-Abelian Kuramoto models}, the term that emphasizes that the underlying manifold is a non-commutative group, see \cite{GZZZW,HKR,Lohe1}. The classical Kuramoto model can be referred to as {\it Abelian Kuramoto model}, a special case for the commutative group $U(1) \equiv SO(2)$ with the group manifold $S^1$.

Alternatively, some authors studied generalized "oscillators" whose states are described by points on spheres $S^{d-1}$. Then the classical Kuramoto model arise as a particular case for $d=2$. Generalized Kuramoto models on spheres have been used in modeling programmable swarms (\cite{O-S}), opinion dynamics (\cite{CLP}), multi-agent systems in geometric consensus theory (\cite{MTG}) and for unsupervised machine learning over multivariate data sets (\cite{CJ}).

In the present paper we derive equations for the order parameter in Kuramoto models on spheres. We compare two classes of such models that have been studied by Tanaka in \cite{Tanaka}. The closed-form equations for the order parameter are derived in the simplest setup with globally-coupled identical oscillators and uniform initial distribution. If the coupling is attractive, population evolves with the time from incoherent state towards the fully synchronized state. Hence, the order parameter monotonically increases from zero to one. However, the pace (velocity) of synchronization process depends on the model and/or the dimension of the sphere.

The derivation of closed-form equations for real order parameters is based on geometric reasoning. For that reason we briefly mention some recent results on relations between Kuramoto models on spheres and hyperbolic geometries in unit balls.

Theoretic study of geometry and global variables in the classical Kuramoto model has been initiated in the seminal paper \cite{WS} of Watanabe and Strogatz. They reported a substitution of variables that reduces the dynamics of $N$ Kuramoto oscillators to the 3-dimensional system of ODE's for global variables. Hence, the evolution of a population of Kuramoto oscillators is restricted to a low-dimensional invariant submanifold. Further investigations of this low-dimensional dynamics evolved into a geometric and group-theoretic study of the Kuramoto model, see \cite{MMS}. The most recent results on relations between Kuramoto models and hyperbolic geometries of the unit disc have been reported in \cite{CEM1,CEM2}.

Low-dimensional dynamics and geometry of generalized Kuramoto models on spheres have been investigated very re\-ce\-ntly in several papers \cite{CGO2,JC,Lohe2,Lohe3,Tanaka}. There are two non-equivalent ways to extend the classical Kuramoto model from $S^1$ to higher-dimensional spheres $S^{d-1}$. Tanaka in \cite{Tanaka} introduced two classes of Kuramoto models on spheres: real models on $S^{d-1}$ and complex models on spheres $S^{2m-1}$ in complex vector spaces. These two classes correspond to two non-equivalent geometries in unit balls. This geometric background has been to a large extent clarified very recently in \cite{LMS}. It has been pointed out that real Kuramoto models generate conformal mappings of unit balls, while complex Kuramoto models are related to analytic automorphisms of unit balls in complex vector spaces. For $d=2$ both mappings reduce to M\" obius transformations of the unit disc, leading to the Poincar\' e model of planar hyperbolic geometry.

Throughout the paper we work under the following assumptions on the initial distribution of oscillators:
\begin{description}
\item[(A1)] The number of oscillators $N$ is infinite and the distribution of oscillators on the sphere at each moment $t>0$ is given by a density function $\rho(t,x), \, t>0, \, x \in S^{d-1}$.

\item[(A2)] The initial distribution of oscillators is uniform on the sphere, i.e. $\rho(0,x) = \frac{1}{P_d}$ for $x \in S^{d-1}$, where $P_d = \frac{2 \pi ^{d/2}}{\Gamma(d/2)}$ is the surface of $S^{d-1}$.
\end{description}
The class of real Kuramoto models on spheres is analyzed in the next Section. We derive equations for the real order parameter, based on recent results about geometry and low-dimensional dynamics in these models. In Section \ref{sec:3} an analogous analysis is conducted for complex Kuramoto models on spheres in even-dimensional spaces. In Section \ref{sec:4} we briefly address quaternionic Kuramoto model that has been introduced in \cite{JC}. Finally, Section \ref{sec:5} contains a brief discussion and outlook.

\section{Evolution of the order parameter in Kuramoto models on spheres}\label{sec:2}
The state of a generalized "oscillator" is given by a unit vector $x$ in the real vector space ${\mathbb R}^d$ and evolves by the following ODE
$$
\dot x = W x,
$$
where $W$ is an anti-symmetric $d \times d$ matrix interpreted as a frequency of the oscillator.

System of coupled oscillators on the sphere $S^{d-1}$ is introduced as follows (\cite{CGO1,Lohe2,Tanaka})
\begin{equation}
\label{realKuramoto}
\dot x_j = W x_j + f - \langle x_j,f \rangle x_j, \quad j=1,\dots,N,
\end{equation}
where $f = f(x_1,\dots,x_N)$ is a global vector-valued coupling function and the notion $\langle \cdot, \cdot \rangle$ stands for the inner product in ${\mathbb R}^d$.

Underline that (\ref{realKuramoto}) is the system of identical oscillators (i.e. all intrinsic frequencies are equal) with global coupling.

It has been shown in \cite{Lipton,LMS} that generalized oscillators in (\ref{realKuramoto}) evolve by the action of the group of conformal mappings of the real vector space ${\mathbb R}^d$. In order to explain that, denote by $G$ the subgroup of conformal mappings in ${\mathbb R}^d$ that preserve the unit ball $B^d$. General map from $G$ can be written as (\cite{Stoll})
\begin{equation}
\label{hyperbolicisometry}
g(x) = R \left( \frac{(-x + |x|^2 a)(1-|a|^2)}{1 - \langle a,x \rangle + |a|^2 |x|^2} + a \right),
\end{equation}
where $a \in B^d$ and $R \in SO(d)$.

Mappings from $G$ are isometries of the unit ball in hyperbolic metric in $B^d$.

\begin{proposition}\label{prop:1} \cite{Lipton,LMS}
Consider a population of coupled generalized oscillators evolving by (\ref{realKuramoto}). Then there exists a one-parametric family $g_t \in G$, such that
$$
x_j(t) = g_t(x_j(0)), \quad \forall t > 0, \quad j=1,\dots,N.
$$
\end{proposition}

Now, consider (\ref{realKuramoto}) in thermodynamic limit and assume that the initial distribution of oscillators is uniform on $S^{d-1}$. Images of the uniform measure on $S^{d-1}$ under conformal mappings are measures whose densities are so-called Poisson kernels $P(x;a)$, with $x \in S^{d-1}$, $a \in B^d$, see \cite{KMcC,Stoll}. For fixed $x \in S^{d-1}$ these kernels are harmonic functions for the hyperbolic Laplace-Beltrami operator on $B^d$.  On the other hand, for fixed $a\in B^d$, we have measures on $S^{d-1}$ that integrated against functions on $S^{d-1}$ yield hyperbolic harmonic functions on $B^d$. In this way we obtain the solution to the Dirichlet problem for the hyperbolic metric on $B^d$.

\begin{proposition}\label{prop:2} \cite{LMS}
Under assumptions (A1) and (A2) the distribution of oscillators at each moment $t \geq 0$ is given by the following probability density function
\begin{equation}
\label{hyperbolicPoisson}
P_{hyp}(x;a(t)) = \left( \frac{1 - |a(t)|^2}{|a(t) - x|^2} \right)^{d-1},
\end{equation}
where  $a(t) \in B^d$, $x \in S^{d-1}$.
\end{proposition}

\begin{remark}
It is obvious from (\ref{hyperbolicisometry}) that $a(t) \in B^d$ is an image of zero, i.e. $a(t) = g_t(0)$.
\end{remark}

\begin{proposition}\label{prop:3} \cite{CGO2,LMS}
Under assumptions (A1) and (A2) vector parameter $a(t)$ in (\ref{hyperbolicPoisson}) satisfies the following vector ODE
\begin{equation}
\label{aODE}
\dot a = W a + \frac{1}{2}(1 + |a|^2)f - \langle a,f \rangle a.
\end{equation}
\end{proposition}

Further, introduce the vector order parameter (centroid) of a population of oscillators evolving by (\ref{realKuramoto})
$$
c(t) = \frac{1}{N} \sum \limits_{i=1}^N x_i
$$
and pass to the thermodynamic limit.

Observe the two points: centroid $c(t)$ is the mean value of the distribution (\ref{hyperbolicPoisson}) and $a(t)$ is the image of zero under the map $g_t$.

\begin{proposition}\label{prop:4} \cite{KMcC,LMS}
Suppose that assumptions (A1) and (A2) are satisfied. Then
\begin{equation}
\label{centroid}
c(t) = \mu_{d-1}(|a(t)|) a(t),
\end{equation}
where the function $\mu_{d-1}(y)$ is given by
\begin{equation}
\label{hypergeometric}
\mu_{d-1}(y) = \frac{1+y^2}{2 y^2} \left[ 1 - \frac{1-y^2}{1+y^2} F \left \{ \frac{1}{2};\frac{d-1}{2};\frac{d+1}{2};- \frac{4 y^2}{(1-y^2)^2} \right \} \right],
\end{equation}
and $F$ denotes the hypergeometric series.
\end{proposition}

\begin{remark}
As we can see, centroid $c$ does not coincide with the image of zero $a$. However, these two points lie on the same radius in $B^d$, i.e. vector $c$ is obtained by multiplying vector $a$ by a scalar multiplier $\mu_{d-1}(|a|)$. The situation is still quite involved, as this scalar multiplier depends both on the dimension $d$ and the modulus of vector $a$.
\end{remark}

For $d=2$ the function $\mu_1(\cdot)$ is the constant one. This means that for $d=2$ points $c(t)$ and $a(t)$ coincide. This case corresponds to the classical Kuramoto model.

Now, focus on the special case of (\ref{realKuramoto}) that is obtained for the specific choice of the coupling function
$$
f(x_1,\dots,x_N) = \frac{K}{N} \sum \limits_{i=1}^N x_i = K c.
$$
This standard choice of $f$ yields the standard real Kuramoto model on $S^{d-1}$
\begin{equation}
\label{realKuramotostandard}
\dot x_j = W x_j + K c - K \langle x_j, c \rangle x_j, \quad j=1,\dots,N.
\end{equation}

Denote $p(t) = |a(t)|$ and $r(t) = |c(t)| = \mu_{d-1}(p(t)) p(t)$. Both, $p(t)$ and $r(t)$ are real numbers, $0 \leq p(t),r(t) \leq 1$. Clearly, $r(t)$ is the real order parameter for system (\ref{realKuramotostandard})

One has $a(t) = p(t) u(t)$, where $u(t)$ is a unit vector on $S^{d-1}$. Substituting this into (\ref{aODE}) one readily obtains ODE for $p(t)$. Putting everything together, we state

\begin{proposition}\label{prop:5}
Consider a population of oscillators evolving by (\ref{realKuramotostandard}) and suppose that assumptions (A1) and (A2) hold. Then
\begin{equation}
\label{realorderODE}
r (t) = \mu_{d-1}(p(t))p(t),
\end{equation}
with $p(t)$ satisfying the following real-order ODE:
\begin{equation}
\label{auxrealorderODE}
\dot p = \frac{K}{2} \mu_{d-1}(p) p (1-p^2).
\end{equation}
\end{proposition}

We have obtained explicit equations (\ref{realorderODE}), (\ref{auxrealorderODE}) for the real order parameter under some pretty restrictive assumptions. Presence of function $\mu_{d-1}(\cdot)$ in the right-hand side of ODE for $p(t)$ implies that the rate of synchronization depends on the dimension. For each specific dimension $d$, the function $\mu_{d-1}(\cdot)$ can be calculated in a closed form.

\begin{remark}
The rate (velocity) of synchronization in real models (\ref{realKuramotostandard}) is determined by functions $\mu_{d-1}(\cdot)$ defined by (\ref{hypergeometric}). However, it can be shown that for all $d \geq 2$ functions $\mu_{d-1}(\cdot)$ satisfy the following properties
\begin{itemize}
\item[a)] $\lim_{y \to 1} \mu_{d-1}(y) = 1$,
\item[b)] $\mu_{d-1}(y) \geq 1$ and
\item[c)] $\frac{d \mu_{d-1}(y)}{d y} > 0$ for $0 < y < 1$.
\end{itemize}

From these properties we draw (pretty much expected) conclusion that the complete synchronization takes place in all dimensions whenever the coupling is attractive, i.e. whenever $K>0$.
\end{remark}

We proceed with examples of real Kuramoto models on spheres $S^1,S^2,S^3$ and $S^4$.

\begin{example}[The classical Kuramoto model] For the classical Kuramoto model $d=2$ and the function $\mu_1(\cdot)$ is equal to the constant one. Hence, real order parameter $r$ satisfies ODE
\begin{equation}
\label{order1}
\dot r = \frac{K}{2}(r - r^3).
\end{equation}
\end{example}

\begin{example}[Model on $S^2$]
For the real Kuramoto model on $S^2$ one evaluate the function $\mu_2(\cdot)$ to obtain
$$
\mu_2(p) = \frac{1+p^2}{2p^2} \left( 1 - \frac{(1-p^2)^2}{2p(1+p^2)} \ln \frac{1+p}{1-p} \right)
$$
and, referring to Proposition \ref{prop:5},
$$
r(t) = \mu_2(p(t)) p(t),
$$
where ODE for $p(t)$ reads
$$
\dot p = \frac{K}{4} \frac{1-p^4}{p} \left( 1 - \frac{(1-p^2)^2}{2p(1+p^2)} \ln \frac{1+p}{1-p} \right).
$$
\end{example}

\begin{figure}[t]
\centering
\includegraphics[width=.49\textwidth]{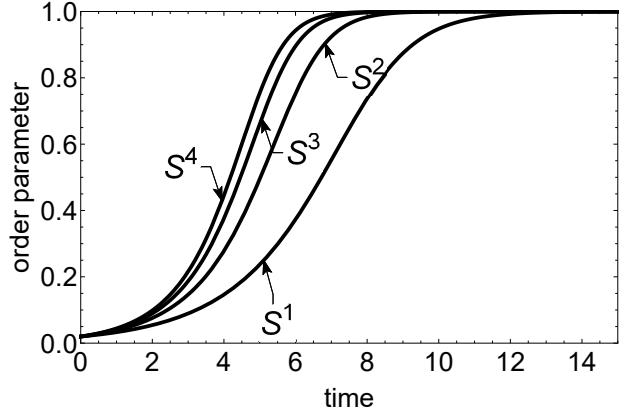}
\caption{\label{fig:1} Evolution of the real order parameters in real Kuramoto models on spheres $S^1,S^2,S^3,S^4$ for coupling strength $K=1$. Initial positions of oscillators are sampled from uniform distributions on spheres. Empirical graphs are obtained by simulating (\ref{realKuramotostandard}) with $N=300$ oscillators. Theoretical graphs are obtained by solving equations for $r(t)$ in examples 1-4. For each dimension these two graphs (empirical and theoretical) coincide.}
\end{figure}

\begin{example}[Model on $S^3$]
One has $\mu_3(p) = \frac{3-p^2}{2}$. Hence, for the model on the 3-sphere
$$
r(t) = \frac{3-p^2(t)}{2} p(t),
$$
with
$$
\dot p = \frac{K}{4}(3p-4p^3+p^5).
$$
\end{example}

\begin{example}[Model on $S^4$]
Computing (\ref{hypergeometric}) for $d=5$ one obtains
$$
\mu_4(p) = \frac{1+p^2}{2p^2} \left( 1 - \frac{3(1-p^2)^2}{8p^2} + \frac{3}{16 p^3} \frac{(1-p^2)^4}{1+p^2} \ln \frac{1+p}{1-p} \right).
$$
Therefore,
$$
r(t) = \mu_4(p(t)) p(t),
$$
where
$$
\dot p = \frac{K}{4} \frac{1-p^4}{p} \left( 1 - \frac{3(1-p^2)^2}{8p^2} + \frac{3}{16 p^3} \frac{(1-p^2)^4}{1+p^2} \ln \frac{1+p}{1-p} \right).
$$
\end{example}

In Figure \ref{fig:1} we depict the real order parameter in models on $S^1,S^2,S^3$ and $S^4$. We have found modulus of centroid $r(t) = |c(t)|$ in two ways: (a) by conducting simulations of (\ref{realKuramotostandard}) for $N=300$ oscillators; and (b) by solving equations for $p(t)$ and $r(t)$ in examples 1-4. In such a way we have obtained two graphs (experimental and theoretical) for each sphere. We obtained perfect fits of simulation results with theoretical predictions. In other words, each graph in Figure \ref{fig:1} is obtained in two different ways. Emphasize that initial positions of oscillators are taken from the uniform distribution in each simulation.

One can see that the pace of synchronization grows with the dimension $d$. For equal values of the coupling strength $K$, synchronization is faster on higher-dimensional spheres.

\begin{figure*}[t]
\centering
\begin{tabular}{@{}cc@{}}
\includegraphics[width=.49\textwidth]{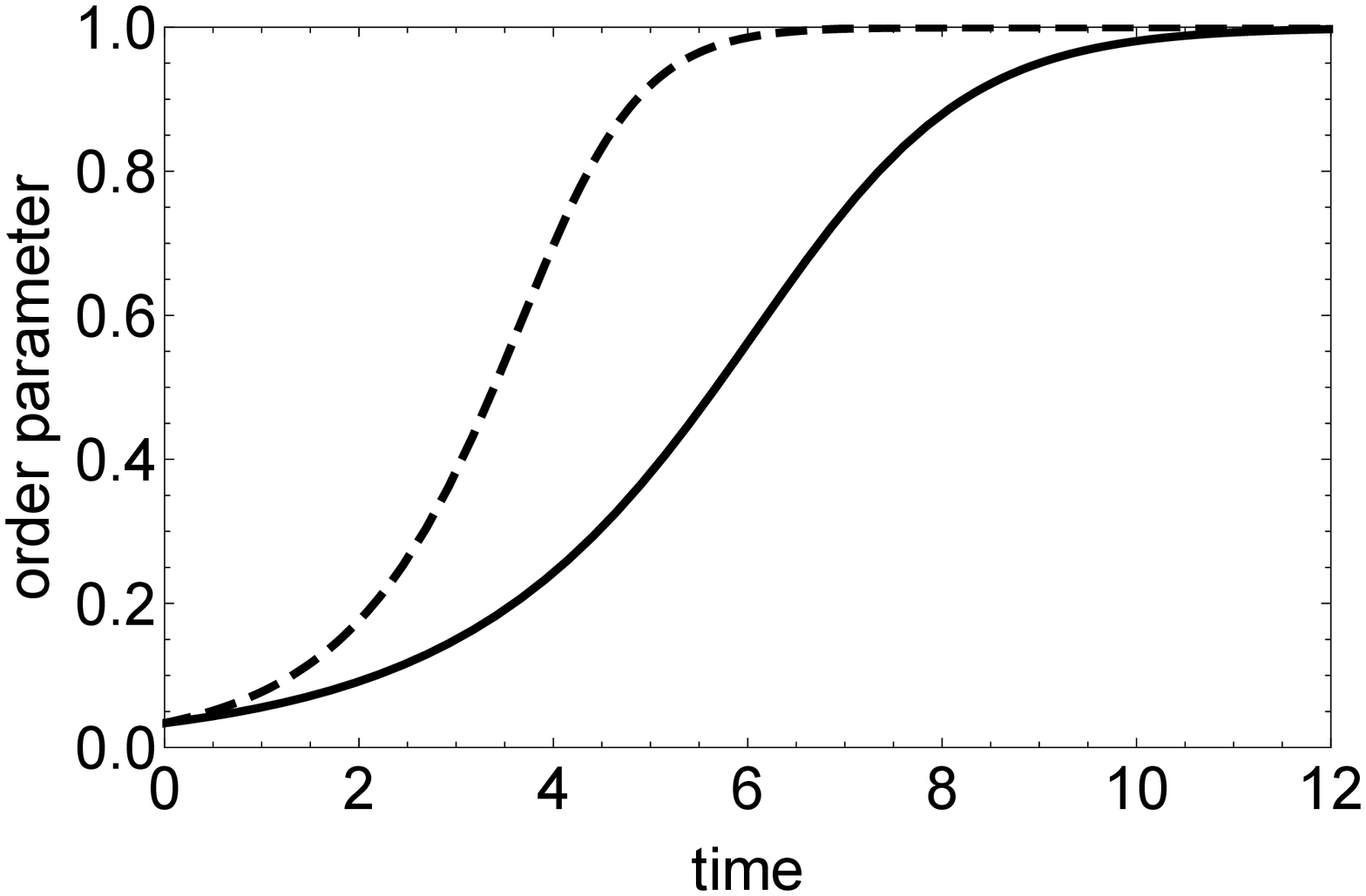} &
\includegraphics[width=.49\textwidth]{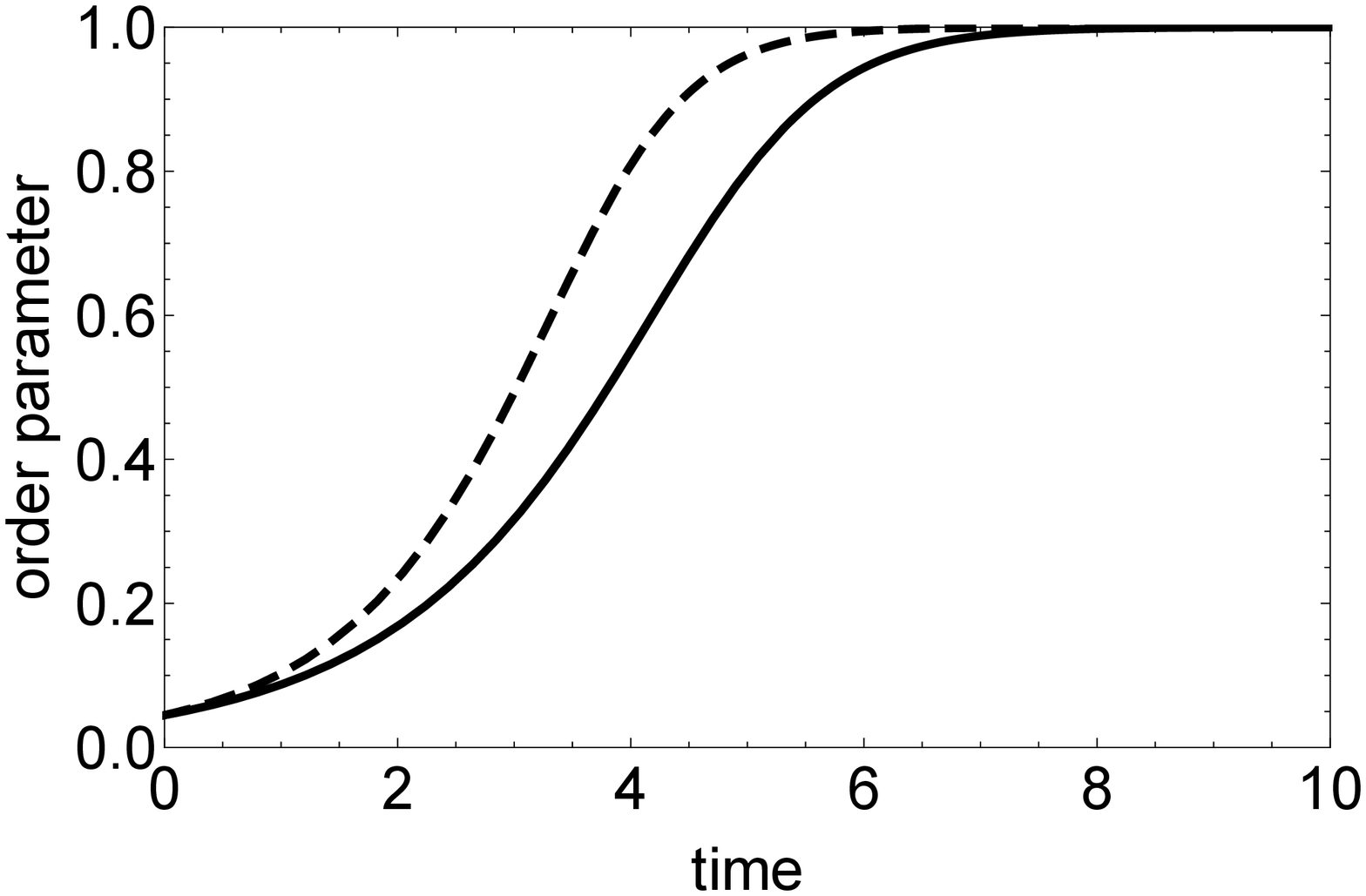}\\
\quad (a)&\quad (b)\\
\includegraphics[width=.49\textwidth]{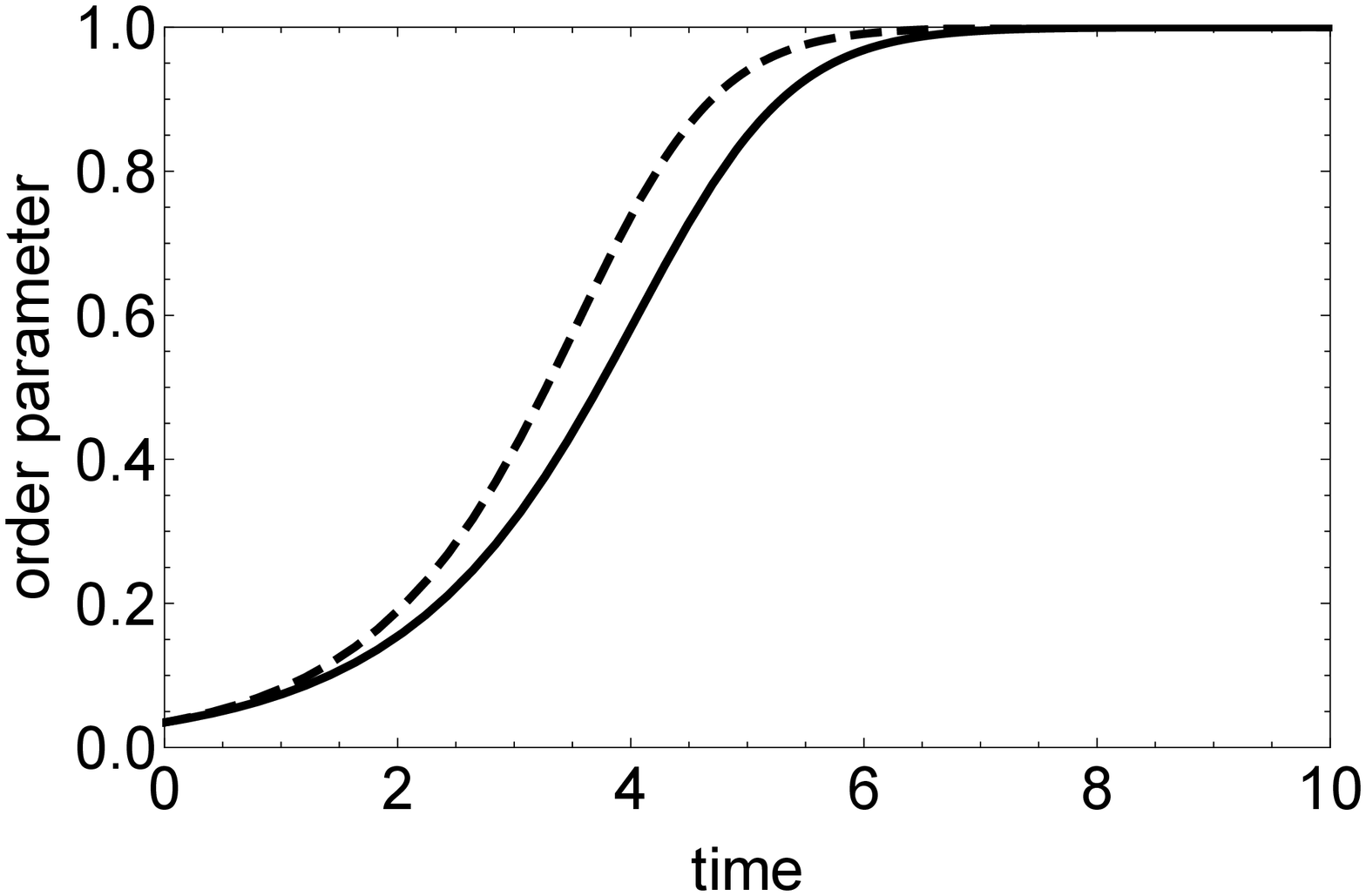} &
\includegraphics[width=.49\textwidth]{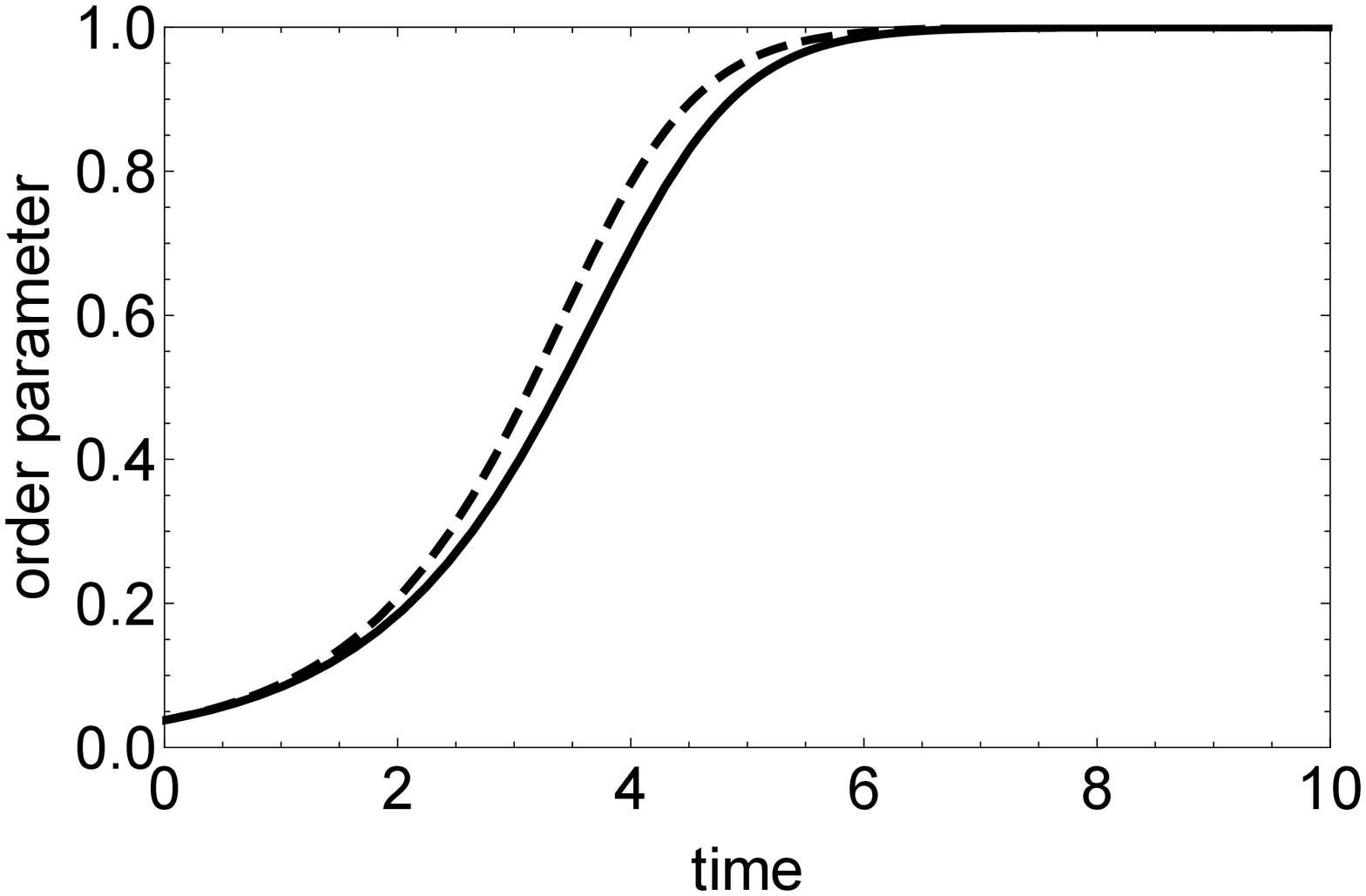}\\
\quad (c)&\quad (d)
\end{tabular}
\caption{\label{fig:2}Actual evolutions of the real order parameters (thick lines) obtained by simulations of (\ref{realKuramotostandard}) with $N=300$ oscillators and coupling strength $K=1$ on spheres: (a) $S^1$; (b) $S^2$; (c) $S^3$ and (d) $S^4$. Initial positions of oscillators are sampled from non-uniform distributions on each sphere. Solutions of equations for $r(t)$ in examples 1-4 are depicted by dashed lines. It is evident that real order parameters do not satisfy equations for $r(t)$.}
\end{figure*}

Figure \ref{fig:2} illustrates significance of assumptions (A1) and (A2). It demonstrates that the real order parameters $r(t)$ do not satisfy equations obtained in examples 1-4 when the initial distribution of oscillators is non-uniform. Actual real order parameters (obtained by simulations) are shown by dashed lines, while solutions of equations 1-4 are shown by thick lines. The initial positions are sampled in the following way: for 150 oscillators positions are sampled from von Mises - Fisher distributions with the concentration parameter $\kappa = 5$ and mean direction $\mu$ and for the remaining 150 oscillators positions are sampled from the same distribution with antipodal mean direction $- \mu$. In such a way we generated $N=300$ points from the balanced non-uniform initial distributions on spheres. From Figure \ref{fig:2} we can see that the synchronization is faster if the initial distribution of oscillators is not uniform.

\section{Order parameter in complex Kuramoto models on spheres in even-dimensional spaces}\label{sec:3}
In the previous Section we have analyzed synchronization in the class of real Kuramoto models on spheres. There is an alternative way to extend the Kuramoto model to higher dimensional spheres. By identifying ${\mathbb C}^m$ with ${\mathbb R}^d$, where $d=2m$, we consider the sphere $S^{2m-1}$ as a subset of ${\mathbb C}^m$, consisting of unit complex vectors.

The complex Kuramoto model on unit sphere in ${\mathbb C}^m$ reads \cite{Tanaka}
\begin{equation}
\label{complexKuramoto}
\dot \xi_j = A \xi_j + p - \langle \xi_j, p \rangle \xi_j, \quad j=1,\dots,N.
\end{equation}
Here, $\xi_j$ are unit vectors in ${\mathbb C}^m$ and $A$ is an anti-Hermitian $m \times m$ complex matrix. This matrix is interpreted as an intrinsic frequency of all oscillators. Coupling is given by the function $p(\xi_1,\dots,\xi_N)$ that takes values in the vector space ${\mathbb C}^m$. From now on $\langle \cdot,\cdot \rangle$ denotes the Hermitian inner product in the complex vector space ${\mathbb C}^m$.

System (\ref{complexKuramoto}) describes a population on the sphere $S^{2m-1}$ in even-dimensional real vector space ${\mathbb R}^d = {\mathbb C}^m$, where $d=2m$. It is easy to check that both models (\ref{complexKuramoto}) and (\ref{realKuramoto}) reduce to the classical Kuramoto model for $d=2m=2$. However, in higher dimensions $d = 2m = 4,6,...$ these two models are not equivalent \cite{LMS}.

Recently, relations of these two classes of models with geometries of unit balls have been explained in \cite{LMS}. It has been shown that generalized oscillators in complex Kuramoto models evolve by the action of the group of isometries of the unit ball in ${\mathbb C}^m$ with the Bergman metric. General orientation-preserving isometry of unit balls with this metric is written as
\begin{equation}
\label{Bergman}
m(\xi) = Q \left( \frac{-\xi + w + \frac{|w|^2 \xi - \langle \xi, w \rangle w}{1 + \sqrt{1-|w|^2}}}{1 - \langle \xi, w \rangle} \right), \quad  |\xi| = 1
\end{equation}
where $w$ is a point in the unit ball and $Q$ is a unitary linear map in ${\mathbb C}^m$.

Denote by $H$ the group of all transformations of the unit ball of the form (\ref{Bergman}).

Infinitesimal generators of $H$ are computed in \cite{LMS}; it is shown that they generate the flow on the complex unit ball of the same form as ODE's (\ref{complexKuramoto}) for oscillators. This form of infinitesimal generators essentially proves the following

\begin{proposition}\label{prop:6} \cite{LMS}
Consider a population of coupled generalized oscillators evolving by (\ref{complexKuramoto}). Then there exists one-parametric family $h_t \in H$, such that
$$
\xi_j(t) = h_t(\xi_j(0)), \quad \forall t > 0, \quad j=1,\dots,N.
$$
\end{proposition}

The next Proposition describes an invariant submanifold on which the system of oscillators evolve if the initial distribution is uniform.

\begin{proposition}\label{prop:7}
Under assumptions (A1) and (A2) the distribution of oscillators at each moment $t \geq 0$ is given by the following probability density function
\begin{equation}
\label{Bergmankernel}
P_{Berg}(\xi;w(t)) = \frac{(1 - |w(t)|^2)^m}{|1 - \langle w(t),\xi \rangle|^{2m}},
\end{equation}
where $|\xi| = 1$ and $|w| < 1$.
\end{proposition}

One can check (see \cite[Section 3.3]{Rudin}) that the mean value of the probability distribution given by (\ref{Bergmankernel}) equals $w$. On the other side, it is obvious from (\ref{Bergman}), that $w(t) = h_t(0)$. Hence, we can state the following

\begin{proposition}\label{prop:8} \cite{Rudin}
Under assumptions (A1) and (A2) the centroid $c(t)$ of the distribution of oscillators at the moment $t>0$ coincides with the image of zero under the action of $h_t$, that is
$$
c(t) = w(t) = h_t(0).
$$
\end{proposition}

Comparison of propositions \ref{prop:4} and \ref{prop:8} unveils a crucial difference between real and complex Kuramoto model on spheres in even-dimensional vector spaces. In complex Kuramoto models centroid coincides with the image of zero under isometries $h_t$ in the Bergman metric in the unit ball. On the other side, in real Kuramoto models the analogous relation between these two points is given by (\ref{centroid}) and equality $c(t) = g_t(0)$ holds only for $d=2$.

By computing infinitesimal generators of $H$ one can verify that parameter $w$ of the distribution (\ref{Bergmankernel}) satisfies the same geometric Riccati equation as oscillators in (\ref{complexKuramoto}), see \cite{LMS}. At the same time, due to Proposition \ref{prop:8}, parameter $w$ is precisely the centroid of the population. Hence, centroid satisfies the same ODE.

\begin{proposition}\label{prop:9}
Under assumptions (A1) and (A2) centroid $c(t)$ of a population satisfies the following complex-vector ODE
\begin{equation}
\label{complexcODE}
\dot c = A c + p - \langle c, p \rangle c.
\end{equation}
\end{proposition}

Now, specify the coupling function $p$ in (\ref{complexKuramoto}) as follows
$$
p(\xi_1,\dots,\xi_N) = \frac{K}{N} \sum \limits_{i=1}^N \xi_i = K c,
$$
where $c$ denotes the centroid of a population. This yields the standard complex Kuramoto model
\begin{equation}
\label{complexKuramotostandard}
\dot \xi_j = A \xi_j + K c - K \langle \xi_j, c \rangle \xi_j, \quad j=1,\dots,N.
\end{equation}

Introduce the real order parameter $r(t) = |c(t)|$, then $c(t) = r(t) u(t)$, where $u(t)$ is a unit vector in ${\mathbb C}^m$. Substitution into (\ref{complexcODE}) yields real-valued ODE for $r(t)$.

\begin{proposition}\label{prop:10}
Consider a population of oscillators evolving by (\ref{complexKuramotostandard}) and suppose that assumptions (A1) and (A2) hold. Then, real order parameter satisfies the following real-valued ODE
\begin{equation}
\label{realorderODEcomplex}
\dot r = K (r - r^3).
\end{equation}
\end{proposition}

\begin{figure}[t]
\centering
\includegraphics[width=.49\textwidth]{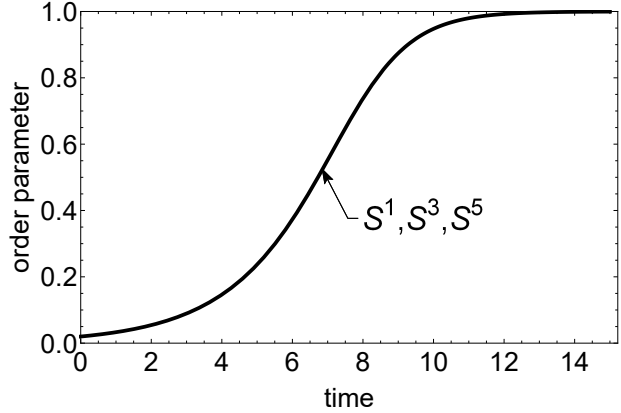}
\caption{\label{fig:3}Evolutions of the real order parameters in complex Kuramoto models on spheres $S^1,S^3$ and $S^5$ for coupling strength $K=1$. Initial positions of oscillators are sampled from the uniform distribution on each sphere. Empirical results are obtained by simulations of (\ref{complexKuramotostandard}) on $S^1,S^3,S^5$ with $N=300$ oscillators. Theoretical result is obtained by solving ODE (\ref{realorderODEcomplex}). All 3 graphs coincide.}
\end{figure}

As we can see, equations for the order parameter in complex Kuramoto models are simpler then those in real models. Moreover, unlike real models, equation for the order parameter is the same in all dimensions.

In Figure \ref{fig:3} we depict graphs of functions $r(t)$ for spheres $S^1,S^3,S^5$. These graphs are obtained from simulations of the system (\ref{complexKuramotostandard}) for $m=1,2,3$, as well as by solving (\ref{realorderODEcomplex}). In all dimensions graphs coincide. Hence, the graph depicted in Figure \ref{fig:3} is obtained in four ways: by conducting simulations (\ref{complexKuramotostandard}) in three dimensions and by solving ODE (\ref{realorderODEcomplex}) with the initial condition $r(0) = 0$.

\section{Quaternionic Kuramoto model}\label{sec:4}
In this Section we briefly mention the quaternionic Kuramoto model on the sphere $S^3$ that has been introduced and studied by the authors in \cite{JC}. Algebra of quaternions provides one possible way to introduce coordinates on $S^3$. Each oscillator is represented by a unit quaternion. The set of unit quaternions is a Lie group with the group manifold $S^3$ and the corresponding Lie algebra is given by the set of "pure" quaternions. (Quaternion $q = q_1 + q_2 \cdot i + q_3 \cdot j + q_4 \cdot k$ is called "pure", if $q_1 = 0$.). An individual oscillator evolves by the following quaternion-valued ODE
$$
\dot q = u q + q v,
$$
where $q(t)$ is the state of an oscillator and $u$ and $v$ are pure quaternions, interpreted as its left and right frequencies, respectively.

Further, the model of coupled oscillators on $S^3$ is written as the system of geometric quaternion-valued Riccati ODE's (\cite{JC})
\begin{equation}
\label{quatKuramoto}
\dot q_j = q_j f q_j + u q_j + q_j v - \bar f, \quad j=1,\dots,N,
\end{equation}
where $f(q_1,\dots,q_N)$ is a quaternion-valued coupling function.

This model is essentially equivalent to the real Kuramoto model (\ref{realKuramoto}) for $d=4$. In order to verify this equivalence it suffices to write (\ref{quatKuramoto}) in real coordinates and make a suitable change of variables in order to relate matrix $W$ with pure quaternion $u$ and $v$.

Hence, the model (\ref{quatKuramoto}) does not bring any qualitative novelty compared to real models. It might however be advantageous to work in quaternionic notation, as it provides an opportunity to use nice geometric properties of unit quaternions, or, equivalently, of $SU(2)$ matrices.

\section{Conclusion}\label{sec:5}
We have derived closed-form equations for evolutions of real order parameters in real and complex Kuramoto models on spheres. Our derivation is based on recent geometric insights into these models. The equations reported here are valid only if the initial distribution of oscillators is uniform on the sphere.

Synchronization in real Kuramoto models is faster in higher dimensions, as illustrated in Figure \ref{fig:1}. On the other hand, in all complex Kuramoto models, order parameters satisfy the same simple ODE (\ref{realorderODEcomplex}) regardless of the dimension of the underlying sphere.

Throughout the paper we have assumed that oscillators are identical and the coupling is global. In such a setup, there is no qualitative difference in the dynamics of order parameters in odd and even dimensions. Underline that in real models with non-identical oscillators there is a sharp qualitative difference in synchronization processes in odd and even dimensions, as recently reported in \cite{CGO1}. It has been shown that in real Kuramoto models on spheres $S^{2k}, \, k=1,2,...$ partial synchronization occurs for an arbitrary weak coupling $K>0$. On the other hand, for models on spheres $S^{2k-1}$ in even-dimensional vector spaces there exists a critical coupling strength $K_c$, such that a partial synchronization takes place only when $K>K_c$; such phase transition is familiar from the classical Kuramoto model.

In the classical Kuramoto model one can derive a simple ODE for the real order parameter even for a population with non-identical oscillators under the condition that their frequencies are sampled from certain prescribed probability distributions on the real line. This result is well known as {\it the Ott-Antonsen reduction}, see \cite{OA}. Analogous results are not available for Kuramoto models on higher-dimensional spheres. Recent geometric insights indicate that the Ott-Antonsen result probably might be extended to real and complex Kuramoto models on spheres $S^{2m-1}$. Clearly, equations for the order parameter in real and complex models would differ significantly, just as they differ in the present paper.


%



\end{document}